\documentclass[%
 reprint,
nofootinbib,
 amsmath,amssymb,
 aps,
prx,
]{revtex4-2}

\pdfoutput=1
\usepackage{graphicx}
\usepackage{tabularx}
\usepackage{rotating}
\usepackage{dcolumn}
\usepackage{bm}
\usepackage[normalem]{ulem} 

\usepackage{amssymb}
\usepackage{pifont}
\usepackage{multirow}
\usepackage{mathrsfs}
\usepackage{mathtools}
\usepackage{makecell}
\usepackage{bbm}
\usepackage{amsmath}
\usepackage{bm}
\usepackage{diagbox} 
\usepackage{siunitx} 

\DeclarePairedDelimiter\ket{\lvert}{\rangle}
\DeclarePairedDelimiterX\braket[2]{\langle}{\rangle}{#1 \delimsize\vert #2}
\usepackage{mathtools}
\usepackage[toc,page]{appendix}

\usepackage{cellspace} %
\usepackage[utf8]{inputenc}

\usepackage[margin=1in]{geometry}
\usepackage{datetime}
\usepackage{xcolor}

\newdateformat{monthdayyeardate}{%
  \monthname[\THEMONTH]~\THEDAY, \THEYEAR}

\usepackage{hyperref}
\usepackage[capitalize]{cleveref} 

\begin{document}

\title{
Mid-circuit ground-state cooling and ancilla readout in the \textit{omg} architecture}
\author{{S. Brudney}, {C. Burns}, {G. J. Gregory}, {E. Ritchie}, {D. J. Wineland}, {D. T. C. Allcock}, and {J. O'Reilly}}
\affiliation{Department of Physics, University of Oregon, Eugene, OR, USA}

\date{\monthdayyeardate\today}
\begin{abstract}
The trapped-ion optical-metastable-ground (\textit{omg}) architecture for quantum processors promises the full functionality of two-species experiments, including sympathetic cooling and non-destructive ancilla readout, without the corresponding hardware overhead. 
We confirm that we can cool a global motional mode of a mixed metastable-ground state Coulomb crystal to the motional ground state via dissipative operations on the ground (\textit{g}) qubit without disturbing coherence of the metastable (\textit{m}) qubit. 
This enables quantum logic spectroscopy to non-destructively readout the state of the \textit{m} qubit using fluorescence detection of the \textit{g} qubit. 
Extensions of these demonstrations to larger system sizes should enable the mitigation of motional heating after ion shuttling and syndrome extraction for quantum error correction, both crucial primitives for future fault-tolerant quantum computers based on trapped ions.
\end{abstract}

\maketitle

\section{Introduction}
Mid-circuit dissipative operations have a wide variety of applications across quantum science and engineering. 
They will be crucial for the operation of large-scale, error corrected quantum computers in the form of state measurement for syndrome extraction~\cite{chiaverini_realization_2004,Fowler2012,bluvstein_logical_2024} and may also prove useful for quantum sensing~\cite{Kessler2014,Pezze2020} and simulation~\cite{FossFeig2021,Tantivasadakarn2023,So2024}. 
Other studies have used mid-circuit preparation and measurement for in-situ noise mitigation~\cite{Singh2023}, more efficient qubit resource allocation~\cite{DeCross2023}, and the study of measurement-induced phase transitions in many-body quantum systems~\cite{Skinner2019,noel_measurement-induced_2022,koh_measurement-induced_2023}. 
Furthermore, trapped ion systems have made use of sympathetic cooling~\cite{NISTsympcoolingfirst}, in which an undisturbed logic ion is cooled via its Coulomb interaction with a co-trapped ion undergoing periods of laser cooling, to enable high-fidelity entangling operations throughout long circuits~\cite{negnevitsky2018,paetznick2024}. 
This technique is especially important in combination with mid-circuit measurement and reset since these operations tend to significantly heat the ions~\cite{Rasmusson2024}.

While the natural replicability of single atoms trapped in ultrahigh vacuum is typically a resource for calibrating and operating large systems, it also makes it more challenging to perform dissipative operations on some atoms while coherently storing information in others. 
This is due to imperfect localization of laser radiation, small spatial separation of co-trapped ions, and the identical transition frequencies of each atom. 
The first two issues can be mitigated to some extent by engineering better optical addressing systems~\cite{motlakunta2024,So2024} or shuttling~\cite{rowe2002transport,gaebler2021,Moses2023,bluvstein_logical_2024} but cannot be completely eliminated.
The last issue has been circumvented using different isotopes~\cite{ballance2015} or elements~\cite{bruzewicz2017,negnevitsky2018,Singh2023}, but this requires significant experimental overhead. 
Several groups have also engineered spectral isolation by using tightly-focused laser beams or Zeeman shifts to reduce crosstalk between neighboring atoms~\cite{Norcia2023,chen2025,ranawat2026sympatheticcoolingYbions,smith2026microwavedrivensamespeciessympatheticcooling}.


Another option is to encode information in a metastable manifold that is spectrally isolated from the fast, cycling transitions typically used for fluorescence-based state detection and laser cooling~\cite{riebe2004,omgBlueprint,Yang2022}. 
This technique has been used to demonstrate ion-photon entanglement~\cite{feng_realization_2024}, mid-circuit measurement, coherent manipulation, and reset of ground-state-encoded qubits while minimally disturbing coherence of the metastable-encoded qubits~\cite{Yang2022,Lis2023}. 
In neutral systems, near-ground state cooling of ancilla qubits in the ground manifold has been shown~\cite{Lis2023}, but there is no clear mechanism for recooling the metastable `data' qubits. 
This may be problematic as heating from recoil and shuttling can degrade entangling gate fidelities and eventually contribute to atom loss~\cite{evered2023}. 
Trapped ions encoding metastable qubits, on the other hand, have been sympathetically Doppler~\cite{Yang2022} and electromagnetically-induced transparency (EIT)~\cite{feng_realization_2024} cooled by co-trapped ground-state ions. 
Sympathetic cooling to near the motional ground state has been demonstrated in multi-species trapped ion systems~\cite{NISTneargroundstatecooling}, but a single-species, mixed-manifold demonstration remained outstanding until this work. 

\begin{figure*}[t]
  \centering
  \includegraphics[width=1.0\linewidth]{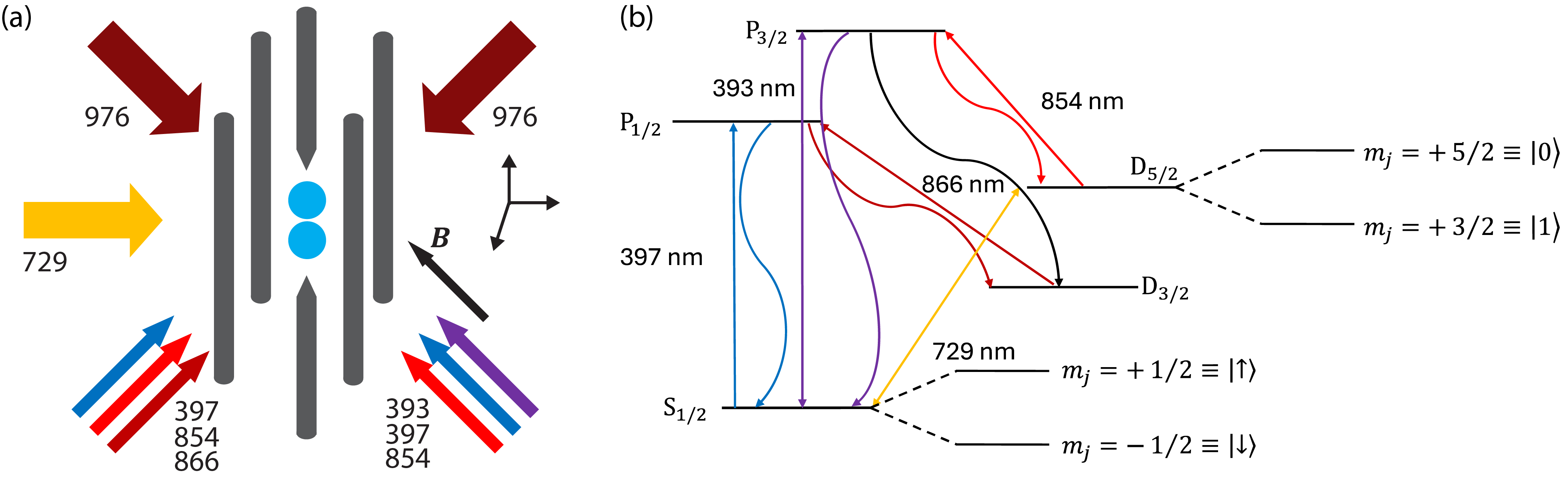}
\caption{Experiment overview. (a) Cartoon of two ions in the trap with directions and wavelengths of incident laser beams in nm. All beam diameters are, in reality, larger than the ion spacing of $\sim6$\,\textmu m.
(b) Level diagram of $^{40}$Ca$^+$ with relevant structure, transition wavelengths, and qubit definitions.}
\label{fig:laser_and_ion_diagrams}
\end{figure*}

Cooling to the motional ground state is typically accomplished using resolved-sideband cooling, in which driving of a phonon-subtracting sideband on a narrow transition is combined with a ``quench" pulse on a broad transition that allows entropy to escape the system via spontaneous emission~\cite{Eschner2003}. 
Usually, the sideband is driven on optical transitions from the ground manifold to a metastable state, which requires a quench that is likely to disturb information encoded elsewhere in the metastable manifold, or on a stimulated Raman transition in the ground state, which requires a second Raman laser system for most trapped atomic species~\cite{Moore2023,Lis2023}. 
Instead, one can drive a sideband from an auxiliary state in a metastable manifold to the ground state and then reject non-ground state motional population via standard fluorescence detection~\cite{eschner1995nulldetectcool,lee2023heraldmotionalground}, also known as `measurement-based cooling.' 
This can be interpreted as erasure conversion for leakage out of the desired motional state, hence the new moniker `erasure-conversion cooling' in recent work~\cite{shaw2025erasure}. 
A related technique has also been dubbed `Raman dark preparation,'~\cite{RDP} but we will use the `measurement-based cooling' terminology in this article.


Furthermore, no experiment yet has shown mid-circuit readout of a metastable qubit via a ground-state ancilla, as will be needed for syndrome extraction in measurement-based quantum error correction. 
We focus on the use of metastable qubits for information processing due to their predicted superior logical performance~\cite{Wu2022,Kang2023,pecorari2025}, which derives from the fact that their fundamental error channels predominantly result in leakage that can be detected and converted to erasure errors. 
Recently, high-fidelity state preparation and measurement (SPAM)~\cite{sotirova2024high} and entangling gates~\cite{gatesLetter,ma2023,scholl2023} as well as minutes-scale coherence times~\cite{Lis2023,Shi2025} have been demonstrated for metastable qubits in trapped ions and neutral atoms.

In the work reported here, we (1) study mid-circuit recooling of the ion crystal after measurement-induced heating and coherent displacement which mimics the residual motional excitation generated by ion shuttling~\cite{sterk2022}; 
(2) integrate a metastable re-shelving technique to demonstrate simple and efficient measurement-based cooling of a global motional mode to $\bar{n}=0.02(1)$; and (3) take advantage of this cooling to perform non-destructive readout of a metastable qubit using fluorescence detection of an identical ancilla ion. 
We use two techniques inspired by quantum logic spectroscopy (QLS), one where we directly drive sidebands to entangle the two ions~\cite{schmidt2005} and another that uses a state-dependent optical dipole force~\cite{Hume2011} and our new measurement-based cooling scheme.



\section{Mid-circuit recooling}

We trap two $^{40}$Ca$^+$ ions in a linear RF Paul trap and use a combination of Doppler cooling, EIT cooling, and optical pumping to prepare each radial motional mode near the ground state $(\bar{n}\approx0.1)$ and each ion's internal state in the S$_{1/2}$ manifold with $m_J=+1/2$~\cite{wineland1985optical}. 
The relevant laser beam and magnetic field geometry, internal structure of the ions, and qubit definitions are shown in \cref{fig:laser_and_ion_diagrams}. 
In order to perform mid-circuit recooling in a two-ion chain, we first prepare one ion as an \textit{m}-qubit in the $\mathrm{D}_{5/2}$ manifold and the other as a \textit{g}-qubit in the $\mathrm{S}_{1/2}$ manifold, which we will refer to as an \textit{mg} crystal. 
To herald this configuration, we apply a global 729\,nm $\pi/2$-pulse on the $\ket{\uparrow}\leftrightarrow\ket{0}$ or $\ket{\uparrow}\leftrightarrow\ket{1}$ transition and perform a fluorescence check (FC) by applying resonant 397 and 866\,nm laser beams and counting photons impinging on a photo-multiplier tube (PMT). 

\begin{figure*}[t]
  \centering
  \includegraphics[width=1.0\linewidth]{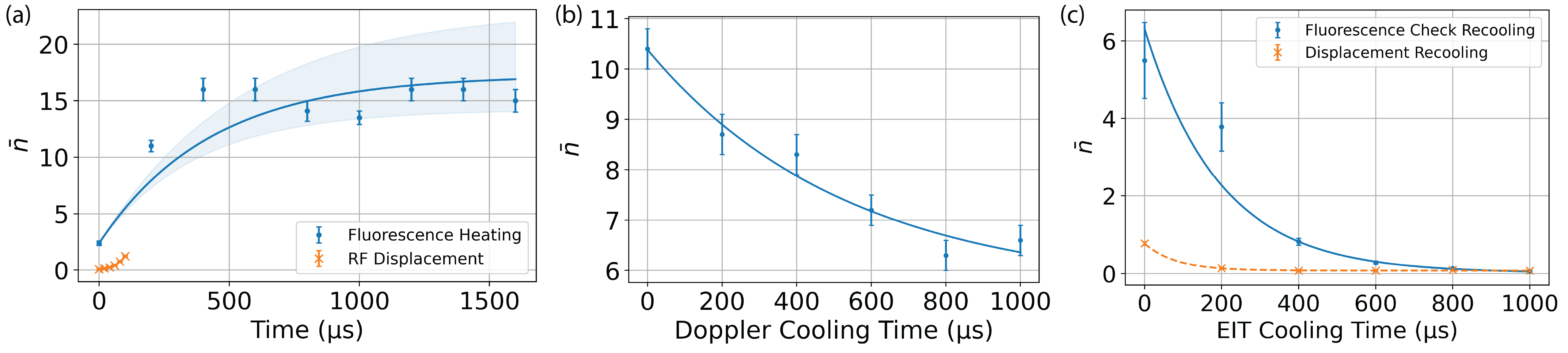}
\caption{(a) Fluorescence heating (solid blue) and RF displacement (dashed orange). 
(b) Doppler cooling after fluorescence heating.
(c) EIT cooling after Doppler cooling (blue) and directly after RF displacement (orange).
}
\label{fig:recooling}
\end{figure*}

Any ion in the \textit{g}-qubit manifold fluoresces during the FC while ions in the \textit{m}-qubit manifold remain dark, so we repeat the 729\,nm pulse and FC until we detect exactly one bright ion. 
Our FC pulses have 1\,ms duration, so the metastable state (natural lifetime of 1.168(9)\,s~\cite{kreuter2005lifetimedstates}) has a small probability ($\lesssim0.1$\%) of decaying in this time frame.
This error does not build up with each successive FC since we re-attempt heralding after each failed check. When we perform another FC after a successful herald to check for consistency, we detect an \textit{mg} crystal $99.81(4)\%$ of the time, limited by D$_{5/2}$ decay during the heralding and checking FCs.

At the end of each experiment, we apply a series of shelving and deshelving pulses and three FCs, each of which can discern between zero, one, and two bright ions, to detect the states of the \textit{g} and \textit{m} qubits and flag any leakage out of the metastable manifold. 
This process, which includes both cabinet shelving~\cite{Benhelm2008} to the $D_{5/2}$ manifold and ``bucket" reshelving via optical pumping to the $D_{3/2}$ manifold~\cite{Vizvary2024} to mitigate imperfections in the 729\,nm pulses, provides measurement fidelities around $99\%$ and is described in detail in the Supplementary Material (SM) Section~\ref{sec:readout}.

 

We studied two sources of motional excitation: fluorescence-induced heating during FCs and RF displacement mimicking residual excitation after shuttling. 
The measured idle heating rate of 1.7(4)\,quanta/s for the 1.86\,MHz center of mass radial (COM) mode in question is negligible on the timescale of these experiments. 
We determined the average Fock state occupation number $\bar{n}$ of this COM mode of the ion crystal by driving Rabi flops on the carrier or phonon-adding motional sideband transition of a two-photon, 976\,nm stimulated Raman transition between the \textit{m}-qubit states. 
The choice of transition was based on the expected range of $\bar{n}$ as the sideband technique can only resolve thermal states up to a few phonons, at which point the two-beam carrier transition, which is less sensitive, becomes a better probe.
The $n$-dependence of the Rabi frequency leads to interference patterns that can be fit to find an underlying Fock state distribution~\cite{Meekhof1996}. 
This approach directly validates the sympathetic nature of the dissipative heating and cooling processes applied to the ground-state ancilla ion. 

During a fluorescence check, the ancilla ion repeatedly absorbs and spontaneously emits photons, which can heat the ion crystal via recoil. 
We perform this operation with the 397\,nm laser beam tuned $-7(2)$\,MHz from resonance and, as shown in \cref{fig:recooling}(a), the crystal approaches the corresponding Doppler temperature of $\bar{n}=17^{+6}_{-3}$ during our standard 1\,ms FC used for two-ion state readout. 
Faster readout can be performed by tuning closer to resonance although this leads to stronger heating~\cite{Rasmusson2024}.
To mimic the coherent motional excitation left after ion shuttling or chain splitting, we applied an RF pulse resonant with the harmonic oscillator COM mode for 100\,\textmu s. 
This resulted in a displaced state with $|\alpha|=1.12(3)$, a value similar to recent demonstrations of shuttling~\cite{clark2023characterization,lancellotti2024low}.

Mid-circuit measurements and shuttling are ubiquitous primitives in quantum computers based on trapped ions~\cite{ransford2025helios}, but the associated motional excitation must be eliminated before high-fidelity laser-based entangling gates can be performed. 
Therefore, after an FC, we tune the 397\,nm laser $-20(2)$\,MHz from resonance and Doppler cool the ion crystal COM mode, observing the time dynamics in \cref{fig:recooling}(b). 
After 1\,ms, the crystal is near the Doppler cooling limit of $\bar{n}=5.6(9)$. 
To cool further, we apply EIT cooling for 1\,ms and approach the calculated limit of $\bar{n}=0.15_{-0.08}^{+0.11}$ (see \cref{fig:recooling}(c)). 
For the displacement, it is sufficient to just apply EIT cooling for 400\,\textmu s. 
In this case, we measure an asymptotic value of $\bar{n}=0.080(6)$.

\begin{figure*}[t]
    \centering
  \includegraphics[width=1.0\linewidth]{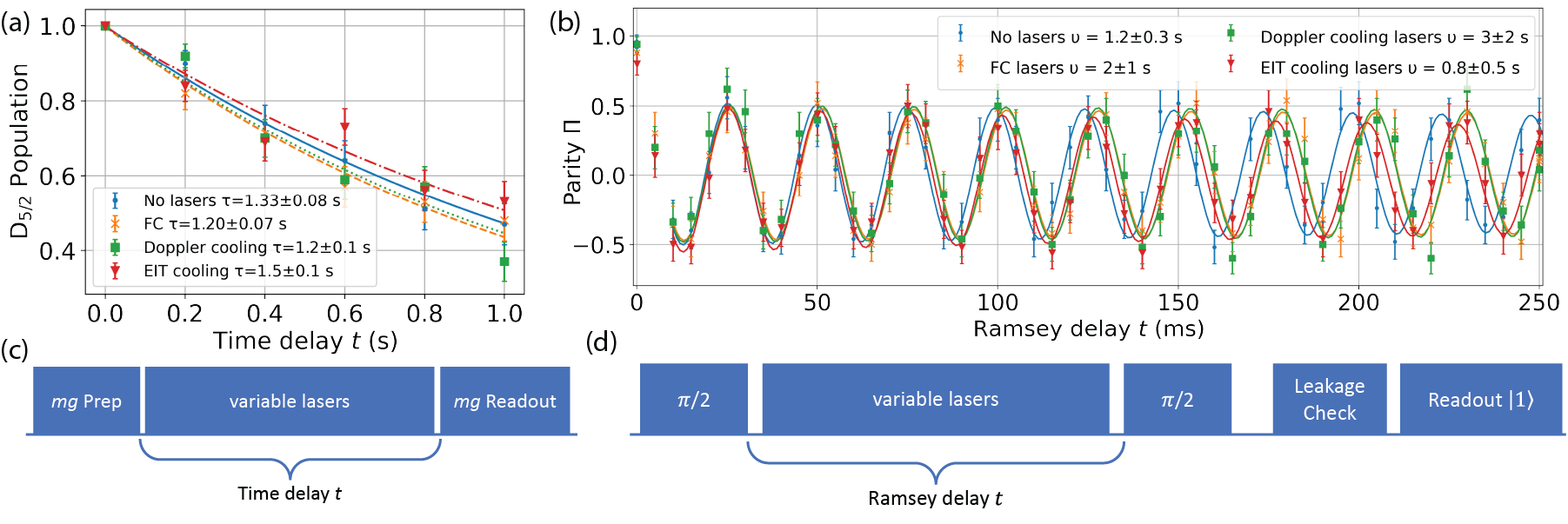}
    \caption{
    Preservation of the \textit{m}-qubit during dissipative operations on the ancilla ion. (a) Lifetime measurement of the D$_{5/2}$ manifold using a single ion while exposed to various sets of laser beams.
    (b) Parity oscillations of the two \textit{m}-qubit protected subspace while ions are exposed to various laser beams. 
    (c) Pulse sequence of lifetime measurement. 
    (d) Pulse sequence of Ramsey experiment to measure parity. 
    }
    \label{fig:nondestruction}
\end{figure*}

Mid-circuit measurement and recooling must preserve the computational \textit{m}-qubits in order to be useful. 
We tested for this by measuring the lifetime $T_1$ of the $\mathrm{D}_{5/2}$ manifold~\cite{kreuter2005lifetimedstates} of a single ion, the spin-echo Ramsey coherence time $T_2$~\cite{cirac1995computecoherence,schneider1998decoherence} of the \textit{m}-qubit in an \textit{mg} crystal, and the coherence of a protected subspace of two \textit{m}-qubits~\cite{chwalla2007protectedsubspace}. 
Each experiment was repeated with the various sets of laser beams required to perform readout and the cooling processes turned on during delay times along with a baseline ``no lasers'' experiment for comparison.

In the D$_{5/2}$ lifetime measurement experiment, a single ion was prepared in the $\ket{0}$ state followed by a variable wait time during which we applied the various sets of laser beams. 
Then, we performed an FC to determine if the \textit{m} population had decayed into the ground state. 
As seen in \cref{fig:nondestruction}(a), none of the beam configurations had a significant effect on the \textit{m}-qubit lifetime.
The measured lifetimes exceed the natural lifetime (1.168(9)\,s~\cite{kreuter2005lifetimedstates}) due to the 393\,nm beam not being fully extinguished, which causes scattering from S$_{1/2}$ back to D$_{5/2}$. 
This prevents some decay events from being detected and biases the lifetime measurement. 

The Ramsey spin echo $T_2$ times with FC, Doppler cooling, and EIT cooling beams on did not significantly differ from the ``No lasers'' experiment, for which we found $T_2=4.0(1)$\,ms. 
This measurement is only sensitive to $\sim300$\,Hz-level qubit frequency fluctuations due to our baseline qubit dephasing, so we also performed a more sensitive measurement using a protected subspace of two \textit{m}-qubits as follows. 

Two ions were prepared in $\ket{\uparrow}$ followed by an RF $\pi$/2-pulse to place each ion in the superposition $\frac{\ket{\uparrow}+\ket{\downarrow}}{\sqrt{2}}$. 
The $\ket{\uparrow\uparrow}$ and $\ket{\downarrow\downarrow}$ states quickly decohere relative to each other and to the protected $\{\ket{\uparrow\downarrow},\ket{\downarrow\uparrow}\}$ subspace. 
The latter is protected from correlated phase noise~\cite{chwalla2007protectedsubspace}, including the significant magnetic field noise in our system and any global Stark shifts from laser beam intensity noise. 
The relative phase in the protected subspace evolves due to the small magnetic field gradient in our trap, which is much more stable than the global average field. 
This produces the parity oscillations seen in~\cref{fig:nondestruction}(b) after a second $\pi$/2-pulse.
We fit the data to a sinusoid with an exponentially-decaying envelope and found that the decay constant $\nu$ was within the uncertainty for all four cases: no lasers $\nu=1.2\pm0.3$\,s, FC lasers $\nu=2\pm1$\,s, Doppler cooling lasers $\nu=3\pm2$\,s, and EIT cooling lasers $\nu=0.8\pm0.5$\,s. 

The results of these experiments show that the implementation of dissipative processes via laser light coupling to the ground-state ion does not significantly effect \textit{m}-qubits encoded in the D$_{5/2}$ manifold of $^{40}$Ca$^+$ ions. 
However, an additional cooling step after Doppler and EIT cooling is needed to reach the motional ground state before performing high-fidelity laser-based entangling gates~\cite{ballance2014thesis}. 


\section{Measurement-based cooling}\label{sec:measurement_based_cooling}

Resolved sideband cooling relies on driving the phonon-subtracting sideband of a narrow transition, most typically between two sublevels of the ground state or from the ground state to a metastable state, and optically pumping back to the original electronic state~\cite{Marzoli1994,monroe1995ramansidebandcooling}. 
For barium and ytterbium, there exist combinations of ground and metastable states where transitions can be conveniently driven by a single stimulated Raman transition laser~\cite{Wang2025,Sotirova2024,qi2026dualtypeYbDuan}. 
For other atomic systems, Raman-based ground state manipulation requires a separate laser system from metastable control due to the $>100$\,THz separation between the relevant dipole transition frequencies.
With this in mind, it would be more convenient in calcium-based trapped ion systems to drive sidebands on the optical transition up to the metastable manifold or, as was recently demonstrated in our group~\cite{gatesLetter}, on a stimulated Raman transition between the \textit{m}-qubit states. 
In both of these cases, the subsequent optical pumping step requires driving the $\mathrm{D}_{5/2}\leftrightarrow\mathrm{P}_{3/2}$ transition (the ``quench'') at 854\,nm. 

\begin{figure*}[t]
    \centering
  \includegraphics[width=1.0\linewidth]{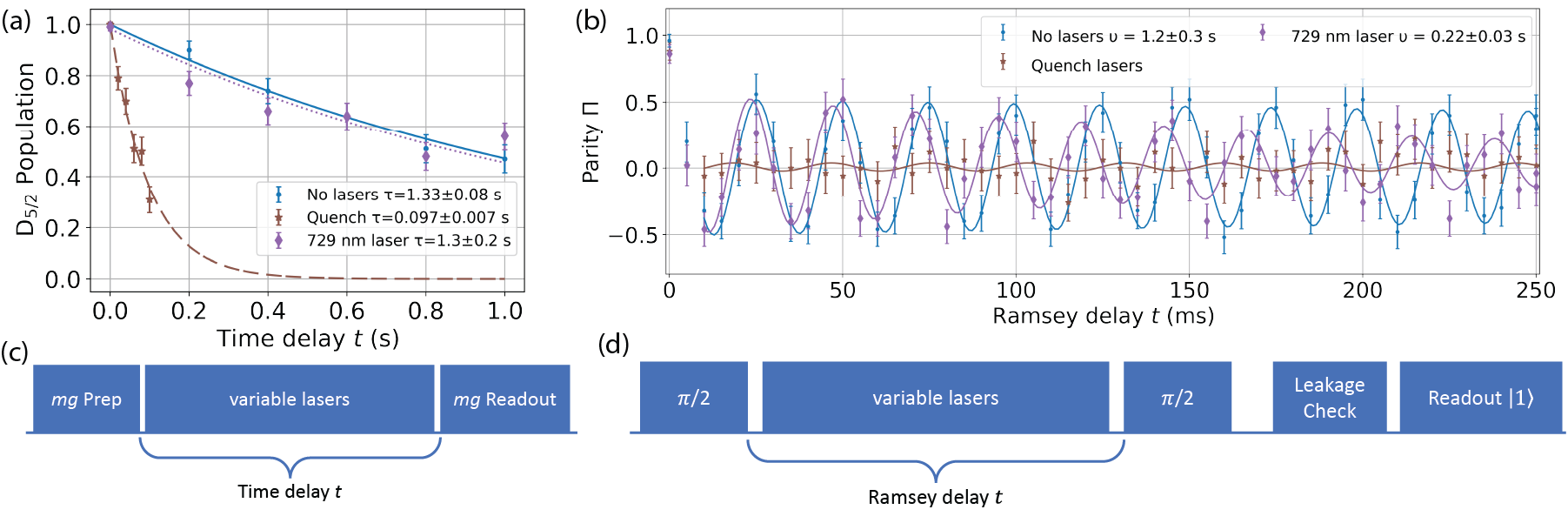}
    \caption{
    Effect of ground-state cooling techniques on the \textit{m}-qubit. (a) Lifetime measurement of the D$_{5/2}$ manifold in an \textit{mg} crystal while ions are exposed to various laser beams.
    (b) Ramsey experiment for two \textit{m}-qubits while ions are exposed to various laser beams. 
    (c) Pulse sequence of lifetime measurement. 
    (d) Pulse sequence of Ramsey experiment to measure parity. 
    }
    \label{fig:nondestruction_quench_729}
\end{figure*}

As shown in \cref{fig:nondestruction_quench_729}, this light induces excessive scattering out of the \textit{m}-qubit and noisy Stark shifts that decohere \textit{m}-qubits even when the quench beam is detuned -20\,GHz from resonance and the polarization is calibrated ($>99\%$ pure $\sigma^+$) to minimize these disturbances. 
This suggests that any sideband cooling technique that quenches D$_{5/2}$ states is incompatible with mid-circuit sympathetic cooling of \textit{m}-qubits. 
As an alternative, we consider driving the phonon-subtracting sideband on the ${\ket{\uparrow_o}\equiv\ket{\textrm{D}_{5/2},m_J=-5/2}\rightarrow\ket{\downarrow}}$ transition. 
This enables sideband cooling mediated by the quadrupole transition and pumping with 397\,nm light, although we instead focus on the measurement-based technique first proposed~\cite{eschner1995nulldetectcool} and demonstrated~\cite{lee2023heraldmotionalground} for trapped ion motional state initialization and more recently adapted to neutral atoms in optical tweezers~\cite{shaw2025erasure}. 

\begin{figure*}[t]
    \centering
  \includegraphics[width=1.0\linewidth]{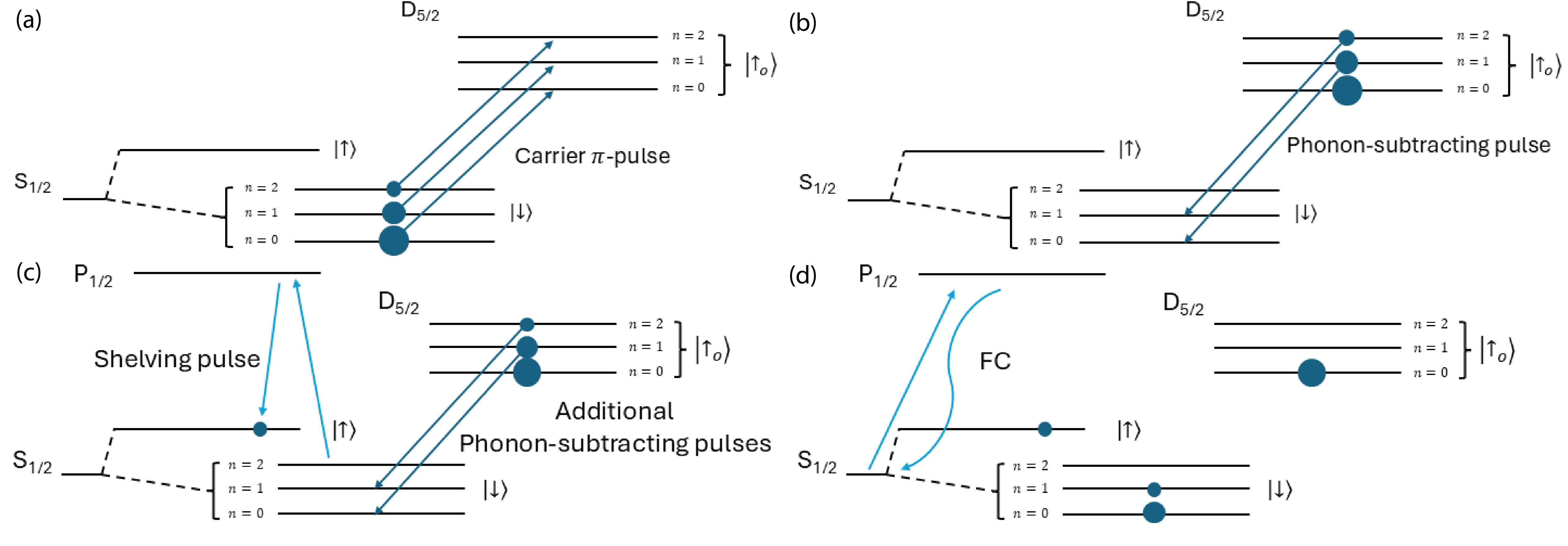}
  \caption{\label{fig:projection_cooling_diagram}
    Diagram of our measurement-based cooling technique. 
    Relative motional populations are represented by size of the circles corresponding to each Fock state. 
    (a) The $g$-qubit is Doppler and EIT cooled, prepared in $\ket{\downarrow}$, and then is shelved with a 729\,nm carrier $\pi$-pulse to $\ket{\uparrow_o}\equiv\ket{\mathrm{D}_{5/2},m=-5/2}$. 
    (b) A 729\,nm pulse on the first phonon-subtracting sideband drives population in motional states $n$\,$>$\,0 from $\ket{\uparrow_o}$ back to $\ket{\downarrow}$. 
    (c) Population in $\ket{\downarrow}$ is pumped into $\ket{\uparrow}$ to allow an additional phonon-subtracting pulse to bring $n$\,$>$\,0 population in $\ket{\uparrow_o}$ that was missed by the previous deshelving pulse down to the ground state. 
    This combination of pulses can be repeated to increase the likelihood that all nonzero $n$ population is deshelved from $\ket{\uparrow_o}$. 
    (d) A fluorescence check is made to detect any population deshelved from D$_{5/2}$. 
    }
\end{figure*}

To perform this measurement-based cooling technique, we first prepare two ions in the \textit{mg} crystal configuration and apply Doppler and EIT cooling for 1\,ms each followed by optically pumping the \textit{g}-qubit to the $\ket{\uparrow}$ state. 
Following the sequence shown in~\cref{fig:projection_cooling_diagram}, an RF $\pi$-pulse transfers the \textit{g}-qubit population to the $\ket{\downarrow}$ state before a 729\,nm laser carrier $\pi$-pulse brings it to $\ket{\uparrow_o}$. 
Next, we attempt to drive population back to the $\ket{\downarrow}$ state via the COM mode's 729\,nm phonon-subtracting sideband.
Ideally, any population in the motional ground state will not be driven back to  S$_{1/2}$ as there is no phonon to subtract. 
Transfer of $n>0$ population is limited by the $n$-dependence of the Rabi frequency~\cite{Meekhof1996,lee2023heraldmotionalground} as well as decoherence during the sideband pulses, so we alternate 729\,nm sideband pulses with optical pumping of all S$_{1/2}$ population to $\ket{\uparrow}$. 
This approach maintains the robustness of Rapid Adiabatic Passage, as employed in Ref.~\cite{lee2023heraldmotionalground}, with less complicated control and better frequency selectivity to avoid driving an undesired transition.
After a prescribed number of pulses, we perform a 500\,\textmu s FC to determine if the ion is in either of the S$_{1/2}$ or D$_{3/2}$ manifolds. 

If the coolant ion is bright, we repeat this pulse sequence starting with sympathetic Doppler cooling to recover from the fluorescence-induced heating. 
Once the coolant ion is dark during an FC, we can proceed with an experiment. 
Importantly, this approach does not require post-selection of data, but rather a pre-selection of prepared states. 
Post-selection requires discarding experiments after they are performed and increases run-times by requiring a larger number of overall shots. 
Pre-selection prevents experiments from continuing during the preparation stage, which reduces run times by not performing experiments that would have been corrupted by imperfect cooling.

For a given number of sideband pulses, we optimize the pulse time by minimizing the height of the phonon-subtracting Raman sideband between the \textit{m}-qubit states. 
For three pulses, we find an optimal length of 170(13)\,\textmu s each and a final average occupation number $\bar{n}=0.02(1)$. 
By fitting a larger range of pulse numbers and initial temperatures to a numerical model described in SM Section~\ref{sec:proj_cooling_simulations}, we find that the cooling is limited by unintended couplings to the $\ket{\downarrow}\leftrightarrow\ket{\uparrow_o}$ and $\ket{\uparrow}\leftrightarrow\ket{\textrm{D}_{5/2},m_J=-1/2}$ carrier transitions. 
The $\ket{\downarrow}\leftrightarrow\ket{\uparrow_o}$ coupling both drives deshelved population back up to D$_{5/2}$, where it contributes to the final temperature measurement, and brings down some of the motional ground-state population, thus reducing efficiency. 
It has a strength of 0.166(4) relative to the on-resonant sideband drive. 
The $\ket{\uparrow}\leftrightarrow\ket{\textrm{D}_{5/2},m_J=-1/2}$ coupling has a strength of 0.019(1) relative to the sideband drive and turns hot population dark, at which point it can again contribute to the final temperature. 
In both cases, the couplings are primarily due to strong servo bumps in our 729\,nm laser spectrum, which could be reduced using feedforward~\cite{li2022active} or cavity filtering~\cite{levine2018high}. 
Given the present values in our system, the model predicts a final temperature of $\bar{n}=0.032(5)$, consistent with our measured value. This could be improved to $\bar{n}=0.001(1)$, even in the presence of dephasing and background heating, by suppressing the servo bumps (see SM Section~\ref{sec:proj_cooling_simulations}).

The probability of passing each FC is fundamentally limited by the motional ground-state population of $\frac{1}{\bar{n}+1}=\frac{1}{0.11(3)+1}=0.90(2)$ after EIT cooling. 
This figure far exceeds the previous demonstration after Doppler cooling of trapped ions, which had a maximum efficiency $\epsilon\approx0.05$~\cite{lee2023heraldmotionalground}, and is greater than the maximum $\epsilon=0.77(1)$ that has been shown for neutral atoms in optical tweezers~\cite{shaw2025erasure}. 
In practice, we achieve $\epsilon=0.79(5)$, consistent with the tweezer result, with most of the discrepancy likely caused by imperfect preparation of the $\ket{\downarrow}$ state and non-unit fidelity of the $\ket{\downarrow}\rightarrow\ket{\uparrow_o}$ carrier $\pi$-pulse. 
These both prevent proper state preparation in $\ket{\uparrow_o}$ at the beginning of the cooling cycle, thus leaving the coolant ion in a bright manifold. 
They could be separated from cooling efficiency in future experiments by applying another FC before the sideband pulses to validate the $\ket{\uparrow_o}$ state preparation step.

Currently, one cooling cycle takes a total of about 3.3\,ms including Doppler and EIT cooling, pumping, three sideband pulses, and the FC. 
On average, for near-unit efficiency, the need to repeat the sequence after a failed attempt increases the time by a factor of $2-\epsilon$. 
The first two steps are roughly minimized for our case but could be made shorter with less aggressive state readout and a tighter trap~\cite{xu20253d}. The sidebands can be made faster by applying more 729\,nm power and the FC can be done in as little as 10\,\textmu s with higher photon collection efficiencies~\cite{myerson2008high,crain2019high}. 
The latter would also allow for a larger detuning during the FC, which in turn would reduce the recooling time. 
Already, the typical timescale is much faster than metastable clock-qubit coherence times~\cite{Shi2025}.

As with the dissipative processes described in the previous section, measurement-based cooling is performed only on the ion designated as the \textit{g}-qubit, which sympathetically cools the ion designated as the \textit{m}-qubit. 
Importantly, results of the experiments using the 729\,nm laser in \cref{fig:nondestruction_quench_729} show that this process does not effect the natural lifetime nor the coherence time of the \textit{m}-qubit and maintains the protected subspace for three orders of magnitude longer than the length of the 729\,nm pulses used ($\upsilon=0.22(3)$\,s versus $t_{729}=170(13)$\,\textmu s). 
This implies that measurement-based mid-circuit recooling is compatible with $^{40}$Ca$^+$ \textit{m}-qubits as logic qubits in the \textit{omg} architecture. 

\section{Non-destructive ancilla readout}

\begin{figure*}
    \centering
  \includegraphics[width=1.0\linewidth]{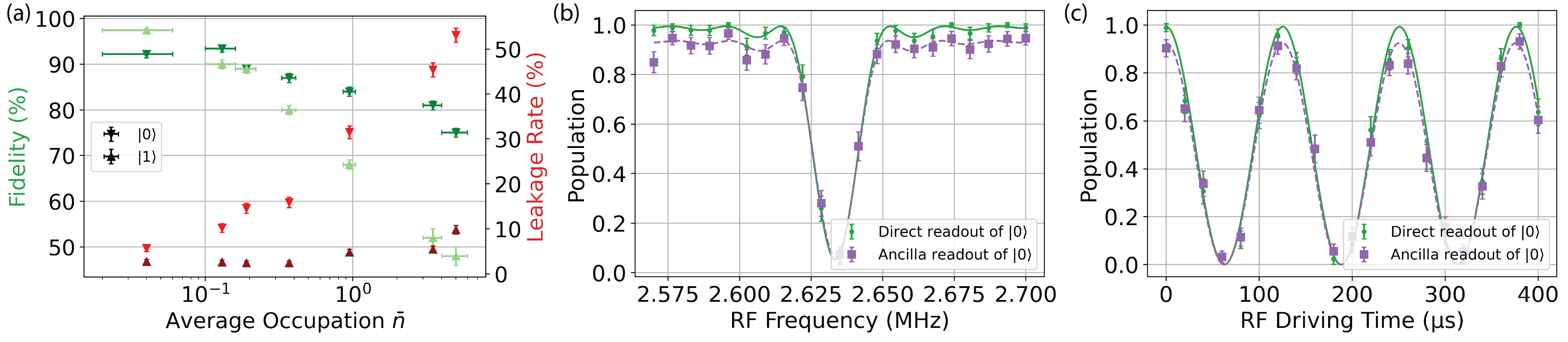}
    \caption{
    (a) Readout fidelity of the \textit{m} state via an ancilla ion in the ground state (left axis) and leakage rate out of the \textit{m} manifold (right axis) versus average occupation as a result of different amounts of cooling. 
    The highest fidelity results occur at the lowest measured $\bar{n}$ value of 0.04(2) achieved using five sideband pulses during measurement-based cooling. 
    (b) RF spectroscopy of the \textit{m}-qubit. 
    The fit gives a center frequency of 2.6335(2)\,MHz for direct \textit{m}-qubit readout and 2.6336(3)\,MHz for ancilla readout. 
    (c) Resonant RF Rabi flopping of the \textit{m}-qubit. 
    We find a Rabi frequency of 250.9(6)\,kHz when fitting $\ket{0}$ population and 250.5(7)\,kHz when fitting $\ket{\uparrow}$ population as the ancilla's proxy for $\ket{0}$ population.
    }
    \label{fig:temp_v_fidelity}
\end{figure*}

Non-destructive readout of a qubit via an ancilla is a crucial primitive for measurement-based quantum error correction~\cite{chiaverini_realization_2004,Fowler2012}. 
Maintaining qubit coherence during readout requires careful isolation between the qubit and ancilla, which is typically achieved using either dual-species operation~\cite{negnevitsky2018,Singh2023} with costly hardware overhead and lower gate fidelities or extra shuttling time~\cite{Moses2023} that slows down feedback cycles. 
\textit{mg} crystals, on the other hand, enable in-situ ancilla readout in a single species system. 
This can be accomplished without strict motional occupation sensitivity using the mixed-type gates demonstrated in Ref.~\cite{Wang2025} or individually-addressed single-type gates combined with Hilbert space shuttling, but here we explore two alternatives that make use of our ground-state cooling capabilities.

\subsection{Quantum Logic Spectroscopy}

First, we perform quantum logic spectroscopy~\cite{schmidt2005} of the \textit{m}-qubit. 
Starting with the ancilla ion in the $\ket{\uparrow_o}$ state, we write the \textit{m}-qubit state to the high-frequency radial COM mode using a 729\,nm carrier $\pi$-pulse that drives $\ket{0}$ population to $\ket{\uparrow}$ followed by a 729\,nm phonon-adding $\pi$-pulse that drives population back to $\ket{0}$. 
Next, a 729\,nm phonon-subtracting $\pi$-pulse drives shelved \textit{g}-qubit population occupying $n=1$ from $\ket{\uparrow_o}$ to $\ket{\downarrow}$. 
Finally, we perform an FC to determine the state of the ancilla ion. 

A bright ancilla indicates that the \textit{m}-qubit is in the $\ket{0}$ state while a dark ancilla corresponds to $\ket{1}$ population. 
Measurement fidelity of the $\ket{1}$ state is limited by residual motional excitation from imperfect cooling, as demonstrated in~\cref{fig:temp_v_fidelity}(a). 
Imperfect contrast of the sideband pulses, which has a thermal contribution as well as a magnetic field-induced dephasing contribution dominant at low temperatures, causes \textit{m}-qubit leakage and $\ket{0}$ readout infidelity. 
To combat the latter effect, we perform bucket deshelving by repeating the phonon-subtracting pulse after pumping any S$_{1/2}$ population to D$_{3/2}$. 
With $\bar{n}=0.04(2)$ we find measurement fidelities of $92.2(8)\%$ and $97.4(5)\%$ and efficiencies of $94.3(7)\%$ and $97.3(5)\%$ for the $\ket{0}$ and $\ket{1}$ states, respectively.

Using this process, we performed spectroscopy and measured Rabi flopping of the \textit{m}-qubit RF transition. 
The fits plotted with the data in \cref{fig:temp_v_fidelity}(b) and (c) give excellent overlap between the \textit{m} and \textit{g}-qubit readouts. 
To confirm that it is non-destructive, we follow the QLS sequence with direct \textit{m}-qubit readout and find an average bit-flip probabilities of 0.7(1)\% 0.4(1)\% for $\ket{0}$ and $\ket{1}$ state preparation, respectively. 
These values are roughly consistent with the generic \textit{m}-qubit SPAM values, implying that the ancilla measurement is in fact non-destructive within the qubit subspace. 
It does cause leakage out of the D$_{5/2}$ manifold in proportion to non-ground state population, but this can be suppressed with technical improvements to improve cooling and sideband driving performance. 
Another interesting possibility is the use of second-order sidebands, although this is even more technically challenging.

\subsection{Spin-dependent force readout}

We can avoid the excessive leakage induced by the previous method by instead writing the \textit{m}-qubit state onto motion using spin-dependent forces. 
This is achieved by interfering two 976\,nm laser beams at the ion crystal to produce an oscillating, state-dependent optical dipole force that is most commonly used to perform two-ion entangling gates~\cite{leibfried2003experimental,gatesLetter}. 
Here, we set the detuning between our two beams equal to the higher radial COM mode frequency to implement the displacement operator $D(\alpha_{0(1)}(t))$ where
\begin{equation}
\alpha_{0(1)}(t)=\frac{\eta\Omega^2_{0(1)}t}{\Delta}
\end{equation}
with the Lamb-Dicke parameter $\eta=0.053$, the single-photon detuning from the D$_{5/2}\leftrightarrow\mathrm{P}_{3/2}$ transition $\Delta=-2\pi\times44$\,THz, and the single-photon Rabi frequency $\Omega_{0(1)}$ depending on the Clebsch-Gordan coefficient for each transition.

By driving each qubit state for variable time and fitting the resulting Fock state distribution to a model for coherent states, we measured displacing rates of $\dot\alpha_{0}=65(3)$\,ms$^{-1}$ and $\dot\alpha_{1}=35(1)$\,ms$^{-1}$. 
Their ratio, $1.86(10)$, is slightly higher than the expected value of $1.67$ based on Clebsch-Gordan differences.

To enable motion-based ancilla readout, population in one of the \textit{m}-qubit states must result in motional excitation while the other state stays in the motional ground state. 
This could be satisfied directly for our qubit encoding using interference of two $\pi$-polarized beams, but geometry restricts us to just one beam containing $\pi$ light. 
We overcome this by applying the spin-dependent force, swapping the \textit{m}-qubit populations with a stimulated Raman pulse, and applying the spin-dependent force (SDF) again with a $\pi$ phase shift relative to the first pulse. 
For example, if we initialize the qubit in the $\ket{1}$ state and the first SDF pulse lasts 20\,\textmu s, then a second SDF time of 11\,\textmu s roughly minimizes the total displacement at $|\alpha_1|=0.355(9)$ while starting in the $\ket{0}$ state gives $|\alpha_0|=0.84(4)$. 
Ideally, we would have $|\alpha_1|=0$ but motional decoherence prevents perfect return to the motional ground state.
Meanwhile, the perturbation in the \textit{m}-qubit subspace can be undone with another population-swapping pulse after the second SDF.

Once information about the \textit{m}-qubit state is written onto the motion, we can again transfer it onto the ancilla ion to complete non-destructive readout. 
Just as before, we drive the phonon-subtracting sideband from $\ket{\uparrow_o}$ to $\ket{\downarrow}$, pump all S$_{1/2}$ population to the metastable D$_{3/2}$ manifold, and drive the sideband again before a final FC. 
Preparation in the $\ket{0}$ state results in a bright detection $51(2)\%$ of the time, which is consistent with its non-ground state population of $e^{-0.70(5)}=0.50(2)$. 
Since the total procedure is non-destructive, we can repeat it including initial cooling and state preparation of the ancilla without disturbing the data qubit. 
The probability of either FC returning a bright ancilla is then $75(1)\%$. 
This figure can be improved by repeating the sequence again or by increasing the displacements, which are currently limited by motional dephasing preventing complete return to the ground state in the `dark' case.

If we instead prepare in the $\ket{1}$ state, the ancilla is dark $88(1)\%$ of the time in the single-shot protocol and dark during both FCs $80(1)\%$ of the time when we perform two readout sequences. 
These figures are consistent with the measured ground-state population of $e^{-0.13(2)}=0.88(2)$ after each set of SDFs. 
Thus, either simple strategy presented above to improve measurement fidelity of the $\ket{0}$ state comes at the cost of worse $\ket{1}$ readout. 
This tradeoff can be broken to some extent by employing adaptive measurement sequences, but efficient, high-fidelity operation will require either better motional coherence ($T_2^*\approx5$\,ms in our system) or selection rules allowing just one state to be driven. 
With the latter, the `bright' state could be driven far from the origin, requiring $|\alpha|>2.15$ for a single-shot fidelity above $0.99$. 
The `dark' state fidelity would then be limited by the initial ground-state cooling probability.

As alluded to above, the latter case could be achieved in our system by changing the running wave polarization from $\sigma^-$ to $\pi$. 
Extending this technique to multiple data ions enables more interesting single-shot readout protocols, including measurement-based entanglement generation that could be useful for distributed quantum computing~\cite{baek2025sdqc} and syndrome readout for stabilization of entangled states~\cite{negnevitsky2018} or full quantum error correction~\cite{chiaverini_realization_2004}.

\section{Summary and Outlook}

This work represents the first demonstration of non-destructive, sympathetic ground-state cooling and ancilla readout compatible with mid-circuit operation in the \textit{omg} or dual-type architecture. 
These are important primitives for maintaining high fidelities in long circuits~\cite{Moses2023} and performing measurement-based quantum error correction~\cite{chiaverini_realization_2004} as well as studies of fundamental phenomena such as measurement-induced phase transitions~\cite{Skinner2019,noel_measurement-induced_2022,koh_measurement-induced_2023}. 
To this end, we have introduced new techniques for the preparation and readout of mixed-type \textit{mg} crystals as well as a simpler way to achieve efficient and robust measurement-based cooling via repumping to an auxiliary state. 
Altogether, this establishes co-trapped ground- and metastable-encoded trapped ion qubits as a versatile experimental platform with much lower overhead than the traditional dual-species approach.

Current results are limited by technical noise, primarily servo bumps in the 729\,nm laser spectrum driving unwanted transitions and magnetic field noise causing dephasing. 
The former can be remedied with feedforward~\cite{li2022active} or cavity filtering~\cite{levine2018high} and the latter can be addressed by driving sidebands faster with more 729\,nm intensity or by installing magnetic shielding~\cite{coherenceImprovement}. 

Eliminating these error sources would enable ground-state population nearing $0.999$ (see SM~\ref{sec:proj_cooling_simulations}), making it competitive with traditional Raman sideband cooling. 
Such performance would be helpful for running laser-based geometric phase entangling gates with high fidelity~\cite{ballance2014thesis}, initializing motional modes for studies of continuous-variable quantum computing~\cite{sutherland2021universal,sinanan2024single,so2025quantum} and sensing~\cite{gilmore2021quantum,valahu2025quantum}, and exploring novel proposals for single-shot many-ion entangling gates~\cite{fang2023realization}. 
It may also be possible to perform true motional erasure conversion of certain two-qubit gate error channels by checking for motional excitation immediately after the gate.

\textit{Acknowledgments} ---
We acknowledge useful discussions with E. Clements. This research is supported in part by the NSF through the Q-SEnSE Quantum Leap Challenge Institute, Award \#2016244 and the US Army Research Office under award W911NF-24-1-0379. The data supporting the figures in this article are available upon reasonable request from J.O.

\bibliography{refs}
\bibliographystyle{apsrev4-1}

\clearpage

\newpage

\section*{Supplementary Material}
\setcounter{section}{0}
\section{Qubit Readout}
\label{sec:readout}

\subsection{\textit{mg} Crystal Readout}

\begin{figure*}[t]
    \centering
  \includegraphics[width=1.0\linewidth]{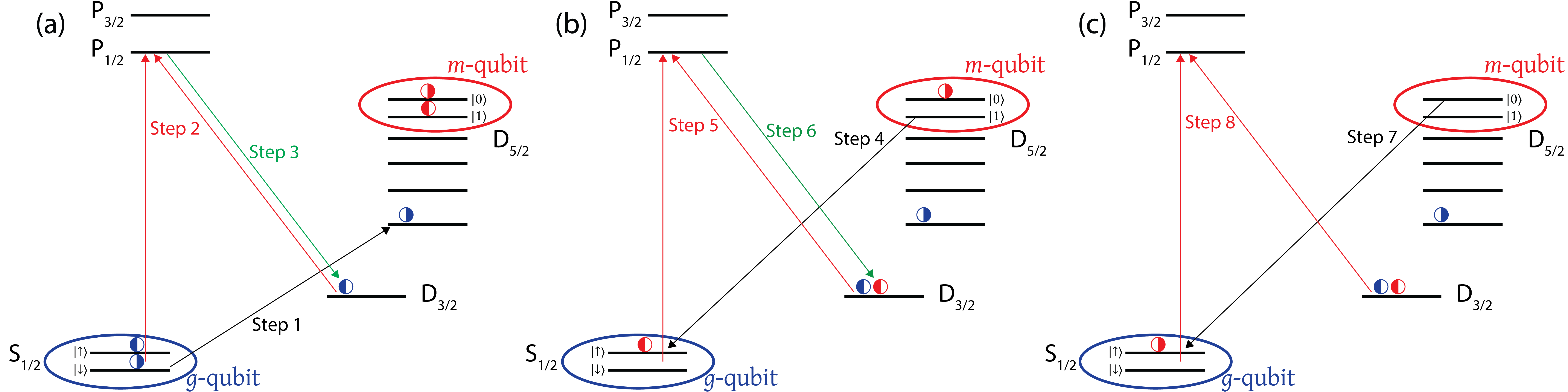}
\caption{Steps for determining population of \textit{g} and \textit{m}-qubits of an \textit{mg} crystal. 
Blue (red) half spheres initially located in S$_{1/2}$ (D$_{5/2}$) indicate where the potential population is moving in each step. 
Arrows indicated population transfer by $\pi$-pulses from appropriate lasers. 
(a) Steps 1--3 determine population of the \textit{g}-qubit. 
(b) Steps 4--6 determine population of the \textit{m}-qubit. 
(c) Steps 7 and 8 deshelve leftover or missed population in D$_{5/2}$.
}
\label{fig:mg_readout}
\end{figure*}

Three fluorescence checks (FCs) along with a series of 729\,nm carrier $\pi$-pulses allows for the determination of each qubit state and conversion of leakage errors into erasures. 
The sequences of numbers of bright ions detected and their corresponding state readout results are listed in \cref{tab:mg_detect}. 

\begin{table}[h] 
    \caption{\label{tab:mg_detect}Corresponding state for a given sequence of bright ion detections in \textit{mg} crystals. 
    }
    \centering
\begin{tabular}{c|c|c
|ll}
\multicolumn{1}{c|}{FC \#1} & \multicolumn{1}{c|}{FC \#2} & \multicolumn{1}{c|}{FC \#3} & \multicolumn{1}{c}{Result} \\
\hline
0 & 0 & 1 & $\ket{\downarrow,0}$ \\
0 & 1 & 1 & $\ket{\downarrow,1}$ \\
1 & 1 & 2 & $\ket{\uparrow,0}$ \\ 
1 & 2 & 2 & $\ket{\uparrow,1}$ \\
0 & 0 & 0 & erasure \\
1 & 1 & 1 & erasure \\
2 & 2 & 2 & erasure \\
\end{tabular}
\end{table}

The basic detection process is as follows, starting with determining the \textit{g}-qubit state, as diagrammed in \cref{fig:mg_readout}(a): 
\begin{enumerate}
    \item $\pi$-pulse on $\ket{\downarrow}\leftrightarrow\ket{\mathrm{D}_{5/2},m=-5/2}$ shelves any population in $\ket{\downarrow}$. 
    \item The first FC is performed. 
        \item S$_{1/2}$ population is optically pumped to D$_{3/2}$ by applying just the 397\,nm beam. 
\end{enumerate}
If one bright ion is detected during the FC, the \textit{g}-qubit was in $\ket{\uparrow}$. 
Otherwise, the 729\,nm $\pi$-pulse shelved $\ket{\downarrow}$ population in $\ket{\mathrm{D}_{5/2},m=-5/2}$ where it will remain for the rest of the full readout sequence. 
Optically pumping any S$_{1/2}$ population into the long-lived metastable D$_{3/2}$ manifold ensures that it does not switch places with the \textit{m}-qubit populations during subsequent 729\,nm pulses. 
This is done by using the 397\,nm laser to drive population to P$_{1/2}$. 
Without the 866\,nm to drive population spontaneously decaying to D$_{3/2}$ back to P$_{1/2}$, population builds up in the D$_{3/2}$ manifold with over 99\% probability in 10\,\textmu s and will remain there until the next FC. 

Next, the \textit{m}-qubit state is determined, as diagrammed in \cref{fig:mg_readout}(b):
\begin{enumerate}
\setcounter{enumi}{3}
    \item $\pi$-pulse on $\ket{\uparrow}\leftrightarrow\ket{\mathrm{D}_{5/2},m=+3/2}$ moves any population in $\ket{1}$ to the S$_{1/2}$ manifold. 
    \item The second FC is performed. 
        \item S$_{1/2}$ population is optically pumped to D$_{3/2}$ by applying just the 397\,nm beam.
\end{enumerate}
If one more bright ion is detected than was in the first FC (recall that population in D$_{3/2}$ will fluoresce), the \textit{m}-qubit was in $\ket{1}$. 
Lastly, population remaining in $\ket{0}$ is detected, as diagrammed in \cref{fig:mg_readout}(c):
\begin{enumerate}
\setcounter{enumi}{6}
    \item $\pi$-pulse on $\ket{\mathrm{S}_{1/2},m=+1/2}\leftrightarrow\ket{\mathrm{D}_{5/2},m=+5/2}$ moves any population in $\ket{1}$ to the S$_{1/2}$ manifold. 
    \item The third FC is performed. 
\end{enumerate}
This final FC ensures that all metastable population was detected. 
The FC results and their corresponding state detections can be found in \cref{tab:mg_detect}.

If, before the readout sequence, the \textit{m}-qubit leaked to $\ket{\downarrow}$, then it will remain dark through all three FCs and the number of bright ions will just depend on the state of \textit{g}-qubit. Similarly, if the \textit{m}-qubit leaked to the D$_{3/2}$ manifold or $\ket{\downarrow}$, it will remain bright through all three FCs. Thus, if all three bright ion detections are identical, we flag an `erasure' error. All sequences not shown in~\cref{tab:mg_detect} are labeled as `error' detections and disregarded.

\subsection{Cabinet Shelving and Bucket Deshelving}\label{sec:cabinet_shelving}

When driving population with a 729\,nm $\pi$-pulse, we rely on the stability of the laser frequency and intensity at the ion to transfer population with high fidelity. 
However, the laser lock is not perfect and the intensity can fluctuate which results in a less than perfect transfer of population. 
Additionally, calibration of the 729\,nm carrier $\pi$-time is imperfect, and finite ion temperatures cause damping due to the carrier Rabi frequency dependence on phonon number. 
To mitigate the resulting non-unity $\pi$-pulse fidelities, we incorporate additional $\pi$-pulses when attempting to shelve and deshelve qubit population in the \textit{mg} crystal readout scheme. 

\begin{table*}[t]
    \caption[SPAM fidelity of each of the four qubit state combinations for an $mg$ crystal.]{\label{tab:mg_spam}
    State preparation and measurement fidelity of each of the four qubit state combinations in an $mg$ crystal.
    Each fidelity is calculated from 10,000 $mg$ crystal preparations. 
    `Leakage' refers to the metastable population decaying before the $m$-qubit population could be determined. 
    All other combinations of detections are categorized as a generic `error.' 
    Fidelity is calculated as the percentage of the correctly prepared states to total prepared states after subtracting out states flagged as a leakage or error. 
    }
    \centering
    \begin{tabular}{l|c|c|c|c|c|c|
        S[table-number-alignment = center]}
    \multicolumn{1}{c|}{\diagbox{Prep}{Result}} & \multicolumn{1}{c|}{$\ket{\uparrow,0}$} & \multicolumn{1}{c|}{$\ket{\uparrow,1}$} & \multicolumn{1}{c|}{$\ket{\downarrow,0}$} & \multicolumn{1}{c|}{$\ket{\downarrow,1}$} & \multicolumn{1}{c|}{Leakage} & \multicolumn{1}{c|}{Error} & \multicolumn{1}{c}{Fidelity (\%)} \\
    \hline
        $\ket{\uparrow,0}$   & 9661 & 63   & 82   & 0    & 183  & 11 & 98.5(1) \\
        $\ket{\uparrow,1}$   & 17   & 9669 & 0    & 77   & 219  & 18 & 99.0(1) \\
        $\ket{\downarrow,0}$ & 107  & 0    & 9546 & 60   & 224  & 63 & 98.3(1) \\
        $\ket{\downarrow,1}$ & 1    & 116  & 15   & 9600 & 211  & 57 & 98.6(1) \\
    \end{tabular}
\end{table*}

Cabinet shelving is used in step 1 of the \textit{mg} readout scheme but was suppressed in the previous section for clarity. 
A single 729\,nm $\pi$-pulse on $\ket{\mathrm{S}_{1/2},m=-1/2}\leftrightarrow\ket{\mathrm{D}_{5/2},m=-5/2}$ sends $\ket{\downarrow}$ population to D$_{5/2}$ so that only $\ket{\uparrow}$ population will fluoresce. 
To increase the probability of shelving all $\ket\downarrow$ population, a second 729\,nm carrier $\pi$-pulse on $\ket{\mathrm{S}_{1/2},m=-1/2}\leftrightarrow\ket{\mathrm{D}_{5/2},m=-3/2}$ can be used to remove any leftover population from the first pulse.  
This process can be repeated as long as there are available states to shelve to. 
For our encoding scheme in $^{40}$Ca$^+$, there are four states available: $m=\{-5/2,-3/2,-1/2,+1/2\}$, with $+3/2$ and $+5/2$ reserved for the \textit{m}-qubit.  
Each pulse gives a population transfer of $>95.3$\%, so two pulses gives a population transfer of $>99.7$\%. 

As a reversal of cabinet shelving, bucket \textit{de}shelving can be used to ensure that all targeted population gets deshelved from D$_{5/2}$. 
After an initial 729\,nm carrier $\pi$-pulse to deshelve $\ket{1}$ population in step 3 or $\ket{0}$ population in step 5, the 397\,nm can again be turned on for 10\,\textmu s to pump S$_{1/2}$ population to D$_{3/2}$, which allows for a second 729\,nm carrier pulse to drive any population not yet driven out of D$_{5/2}$. 
Once again, two 729\,nm carrier pulses each with $>95.3$\% probability of driving population makes the probability of properly deshelving \textit{m}-qubit population to be $>99.7$\%. 

\subsection{\textit{mg} Crystal SPAM Fidelity}\label{sec:mg_spam}

To characterize our state preparation and measurement (SPAM) fidelity for $mg$ crystals, each of the four combinations of the $m$ and $g$-qubit states were prepared 10,000 times and we performed the $mg$ readout scheme described above. 
To prepare the $\ket{\downarrow}$ state, an RF $\pi$-pulse on the $g$-qubit is performed after heralding an $mg$ crystal. 
Each state is prepared and measured with a fidelity of over 98\% after subtracting error and erasure events (\cref{tab:mg_spam}). 

\section{Measurement-based Cooling Simulations}\label{sec:proj_cooling_simulations}
Simulations of our measurement-based cooling protocol were performed via numerical time-dependent Lindbladian evolution using the QuantumOptics.jl package \cite{kramer2018quantumoptics}. Pulses of 729\,nm radiation coupling the $S_{1/2}$ and $D_{5/2}$ manifolds were simulated with a Hamiltonian consisting of a resonant term and, in the case of the sideband pulses, two smaller off-resonant coupling terms. Without considering off-resonant coupling terms, the pulse Hamiltonian is

\begin{equation}
    H_\text{pulse}/\hbar=\frac{\Omega}{2}\left(\sigma_{+}e^{i\eta(a+a^\dagger)}e^{-i\Delta t}+\text{h.c.}\right)
\end{equation}
where spin matrices are defined with respect to $\ket{\uparrow_o}$ and $\ket{\downarrow}$, $\Omega$ is the transition Rabi frequency, $\omega$ is the trap frequency, $\Delta$ is the detuning from the carrier, and $\eta$ is the Lamb-Dicke parameter. Our measurement-based cooling scheme relies on carrier and first-order blue sideband pulses, which correspond to $\Delta=0$ and $\Delta=\omega$, respectively. When $\eta^2(2n+1)\ll 1$ the following time-independent results hold under the rotating-wave approximation \cite{wineland1998experimentalthebible}:

\begin{equation}
    H_\text{carrier}=\frac{\Omega}{2}\sigma_x,\quad H_\text{BSB}=\frac{\Omega\eta}{2}\left(\sigma_-a+\sigma_+a^\dagger\right).
\end{equation}

\begin{figure}
    \centering
    \includegraphics[width=0.45\textwidth]{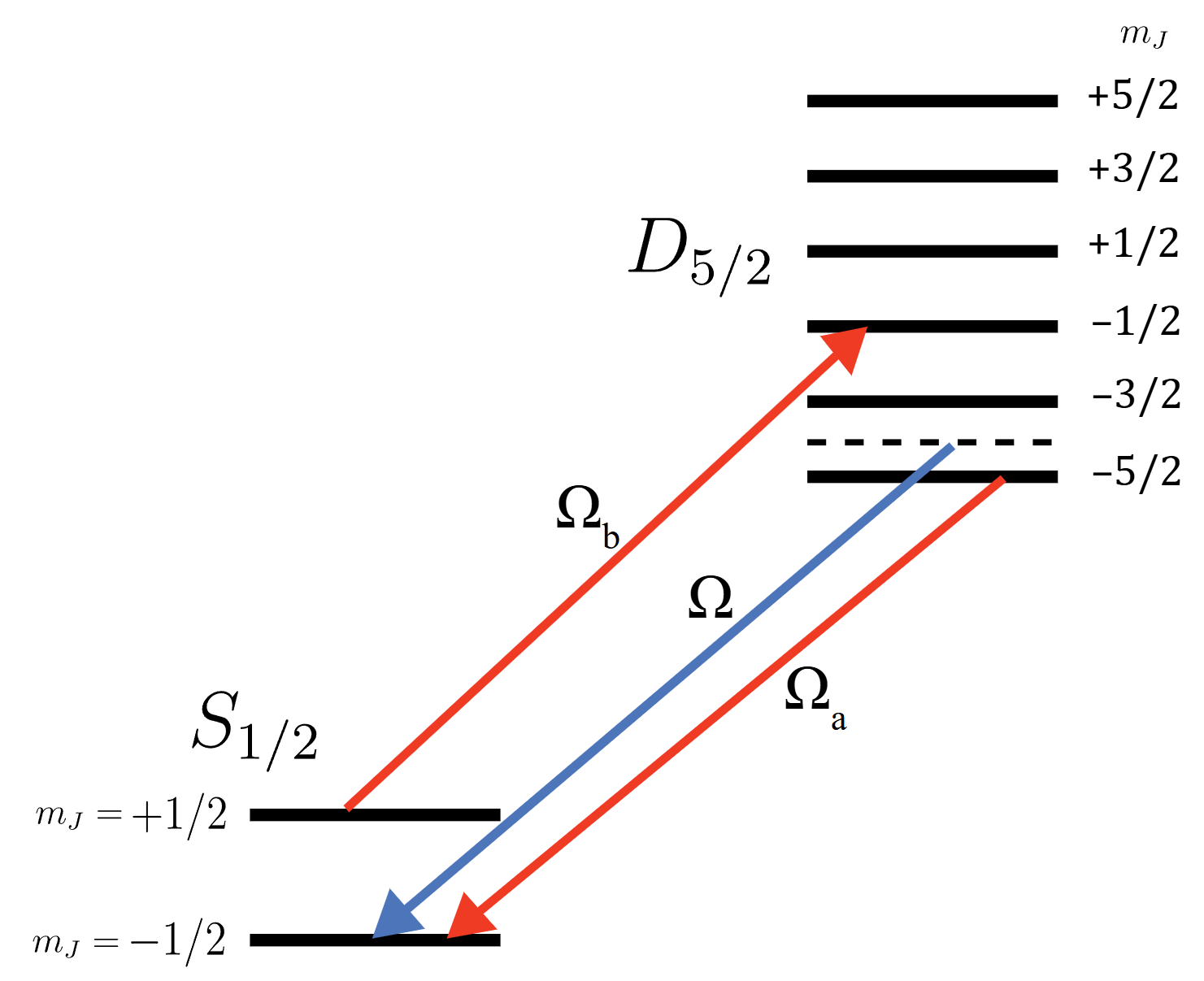}
    \caption{Schematic illustration of cooling sideband and error processes (a) and (b). Process (a) removes cold population from the cold manifold and process (b) moves hot population back into the cold manifold.}
    \label{fig:sup_off_resonant_errors_diagram}
\end{figure}

Our model of the sideband pulses includes additional terms that incorporate off-resonant driving due to classical laser noise resonant with undesired transitions. The two most significant couplings are carrier transitions between (a) the $S_{1/2},m_J=-1/2$ and $D_{5/2},m_J=-5/2$ states and (b) the $S_{1/2},m_J=+1/2$ and $D_{5/2},m_J=-1/2$ states (see Fig.~\ref{fig:sup_off_resonant_errors_diagram}). We label their effective couplings as $\Omega_a$ and $\Omega_b$, respectively, and their Hamiltonian terms are $H_a=\frac{\Omega_a}{2}\sigma_x^{(a)}$ and $H_b=\frac{\Omega_b}{2}\sigma_x^{(b)}$ because they drive carrier transitions. Under reasonable experimental circumstances it is expected that $\Omega_a<\Omega$ and $\Omega_b<\Omega$ so the dominant coupling is on the blue sideband. $\Omega_a$ and $\Omega_b$ were fitted to calibration experiments described below and found to be $\Omega_a=2\pi\times 470(20)\;\text{Hz}$ and $\Omega_b=2\pi\times 53(3)\;\text{Hz}$. Errors were estimated using Monte-Carlo uncertainy estimation, where  the cooling sideband Rabi frequency and initial temperature were sampled according to their measured values of $\Omega=\pi /170(13)\;\mu\text{s}$ and $\bar{n}_0=0.11(3)$

Calibration data was taken by driving a phonon-subtracting sideband of a two-photon Raman transition after a measurement-based cooling sequence with variable pulse times and number. After EIT cooling, the electronic population resides in $D_{5/2}$ and non-ground state motional population will primarily be in the $n=1$ Fock state. As a proxy for the temperature, the 976\,nm Raman red sideband \cite{gatesLetter} is driven for the $n=1\rightarrow n=0$ $\pi$-time to maximize population transfer of the first motionally excited state to the $m_J=+3/2$ state in $D_{5/2}$. Then, population with $m_J<+5/2$ within $D_{5/2}$ is optically pumped to $S_{1/2}$ and we perform an FC. The fluorescence probability measured as a function of the number of cooling pulses and each cooling pulse's duration is sufficient to numerically estimate the couplings $\Omega_a$ and $\Omega_b$ since their effect becomes more pronounced when the 729\,nm beam is on for longer.


In addition, dissipative effects were simulated during 729\,nm laser cooling pulses and during optical pumping to $\ket{\uparrow}$ between cooling pulses with the Lindblad master equation. Motional dephasing was simulated with the dissipator $L_\text{m}=(T_{2,\text{m}})^{-1/2}a^\dagger a$ where $T_{2,\text{m}}=4$ ms. Spin dephasing from magnetic field fluctuations was simulated with $L_\text{s}=(T_{2,\text{s}}^*)^{-1/2}\left(g_SJ_z^{(S)}\oplus g_DJ_z^{(D)}\right)$ where $T_{2,\text{s}}^*=1$ ms and $g_S = 2$ and $g_D = 1.2$ are the Land\'e g-factors of the $S_{1/2}$ and $D_{5/2}$ manifolds, respectively. Lastly, a heating term $L_\text{h}=\sqrt{\dot{\bar{n}}}a^\dagger$, with $\dot{\bar{n}}=1$ quanta/s, is included.

\begin{figure}[ht]
    \centering
    \includegraphics[width=0.45\textwidth]{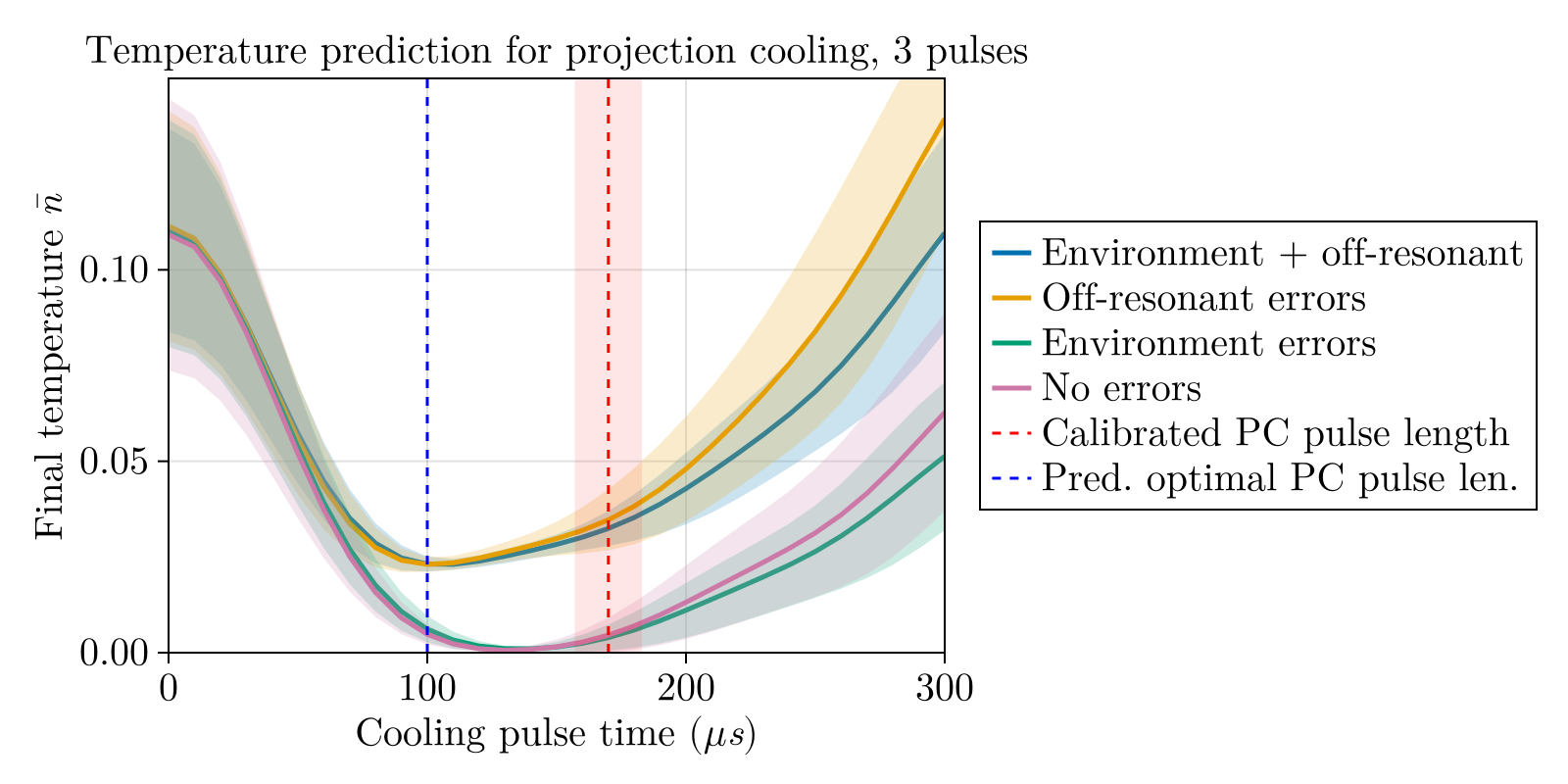}
    \caption{Simulated final average mode occupation $\bar{n}$ as a function of 729 nm sideband cooling pulse time for a three-pulse measurement-based cooling sequence. Off-resonant carrier driving errors contribute most substantively to the achievable temperature lower bound. With only  environmental errors considered, the relevant limit approaches the off-resonant carrier cooling limit of $(\Omega/2\omega)^2\sim 10^{-4}$ \cite{wineland1998experimentalthebible} set by the effective linewidth of the 729\,nm transition. Off-resonant couplings raise the temperature above the ground state to $\bar{n}=0.032(5)$ given an initial temperature of $\bar{n}_0=0.11(3)$ with three $170\;\mu \text{s}$ pulses.}
    \label{fig:sup_3p_temperature_predictions}
\end{figure}

To compare the model with the final measured temperatures cited in Section ~\ref{sec:measurement_based_cooling} and to characterize the impact of different sources of cooling inefficiencies, we simulated the final temperature as a function of the duration of the cooling pulse and plot the results for three applied cooling pulses in Figure \ref{fig:sup_3p_temperature_predictions}. We found that off-resonant couplings due to laser frequency noise have an outsized impact on final temperatures relative to sources of environmental decoherence such as dephasing of the cooling ion's spin or motional degrees of freedom. When cooling pulses last $170\,\mu$s, ideal pulses with and without environmental decoherence achieve temperatures of $\bar{n}=0.004(4)$, which increases by an order of magnitude to $\bar{n}=0.032(5)$ when including off-resonant couplings. The purity of the laser spectrum is therefore the principal technical limitation of measurement-based cooling in our experiment. We also note that improving the experimentally achieved temperature from $\bar{n}=0.032(5)$ to $\bar{n}=0.023(2)$ might be achieved by decreasing the cooling pulse duration to $100\,\mu s$, which the model predicts is optimal in the presence of our off-resonant couplings. Without off-resonant couplings due to laser noise, the model predicts that both with and without heating / dephasing, an optimal $\bar{n}=0.001(1)$ can be achieved after three $140\,\mu$s pulses. This figure roughly aligns with the theoretical minimum temperature achievable in the presence of a heating rate of 1 quanta/s experienced throughout the cooling process.

\begin{figure*}
    \begin{minipage}{0.98\textwidth}
        \centering
        \includegraphics[width=\linewidth]{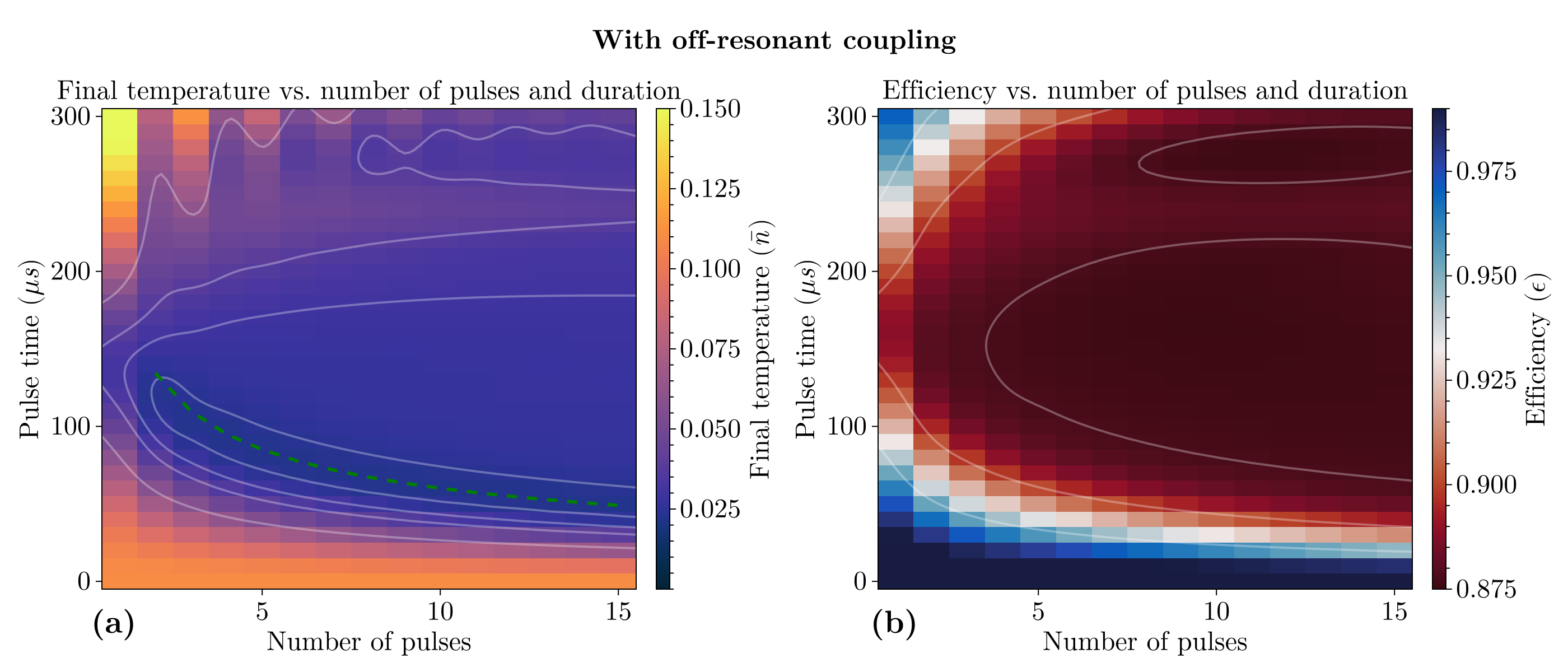}
        \includegraphics[width=\linewidth]{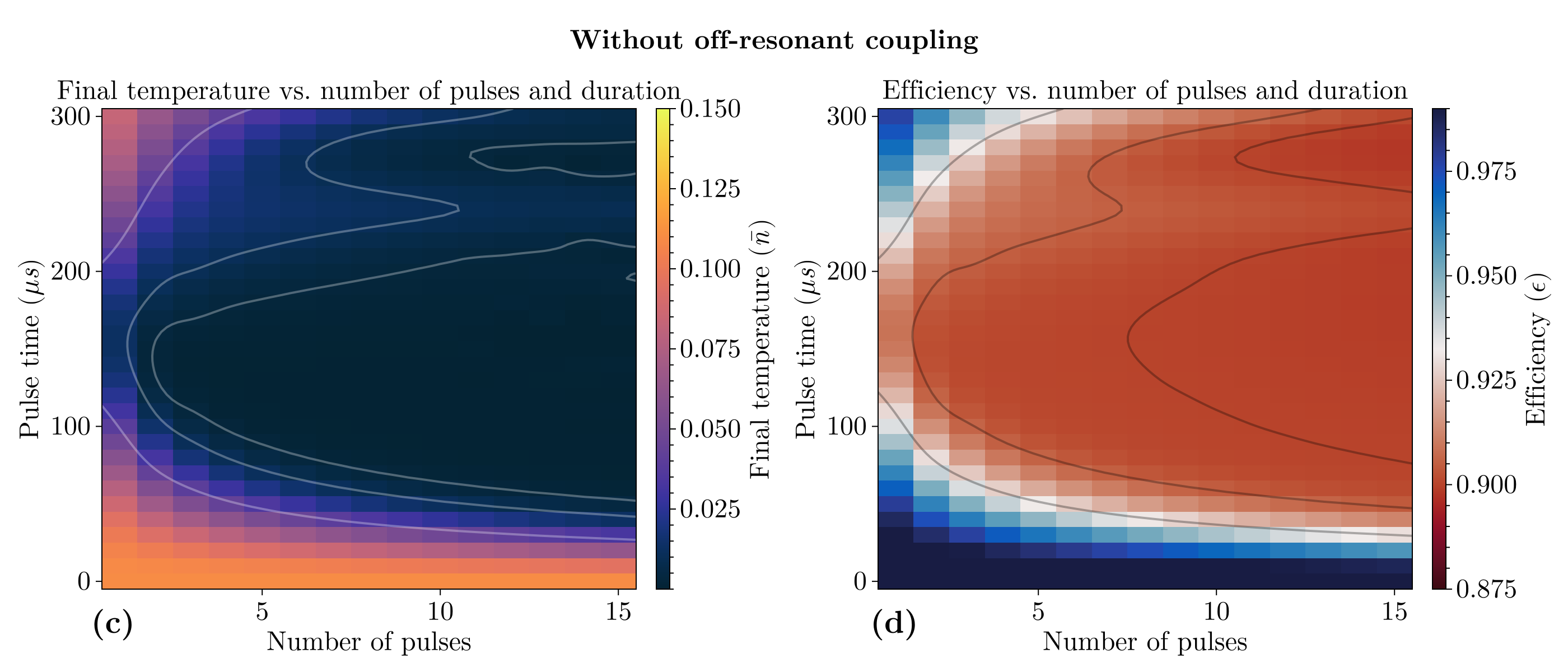}
        \caption{Heatmaps depicting (a/c) the temperature $\bar{n}$ and (b/d) the efficiency $\epsilon$ as a function of number of cooling pulses and the length of each cooling pulse. All four simulations were performed with initial temperature $\bar{n}_0=0.11(3)$ and a sideband $\pi$-time of $170\,\mu s$. (a) and (b) incorporate off-resonant carrier coupling while (c) and (d) simulate cooling pulses without off-resonant coupling. The green dashed line in (a) follows the $N^{-1/2}$ scaling of the optimal cooling pulse time.}
        \label{fig:sup_temperature_efficiency_heatmaps}
    \end{minipage}
\end{figure*}

We also use this model to study whether additional cooling pulses might bring about a further improvement in temperature. We find that optimal pulse time scales with the number of pulses like $N^{-1/2}$ (green dashed line in Figure \ref{fig:sup_temperature_efficiency_heatmaps}). That is, there exists some characteristic $T^\star$ such that $t^\star=T^\star/\sqrt{N}$ where $N$ is the number of pulses and $t^\star$ is the optimal time for each \textit{individual} cooling pulse to be applied. For our setup, $T^\star\approx 190\,\mu s$. In the low-temperature limit, most of the hot population is singly-excited ($n=1$). Each successive cooling pulse reduces the excited-state population by $\cos^2(\Omega t/2)\approx 1-(\Omega t/2)^2$, so the total reduction scales as $(1-(\Omega t/2)^2)^N\approx1-N(\Omega t/2)^2$ since cooled population is shelved after each pulse and is unaffected by subsequent pulses. Rearranging for $t$ we find $t\propto \frac{1}{\Omega \sqrt{N}}$, which shows that for some fixed reduction in excited-state population we expect $t\sim N^{-1/2}$. Along $t=\kappa/\sqrt{N}$ lines within the small-$t$ regime, our model predicts minimal temperature variation across number of pulses. For instance, picking $N=5$, $t^\star\approx 85\,\mu s$ and the predicted temperature is $\bar{n}=0.023(2)$. Taking $N=15$, where $t^\star\approx 50\,\mu s$, the model predicts an identical final temperature of $\bar{n}=0.023(2)$.

\end{document}